\documentclass[prb,twocolumn]{revtex4}
\usepackage{amsmath,amssymb,graphicx}

\newcommand{\md}{{\rm d}}

\newcommand{\rms}{{\rm r.m.s.}}
\newcommand{\hlf}{{\textstyle\frac{1}{2}}}

\begin{document}

\title{\bf Dynamics and performance of clock pendulums}

\author{Peter Hoyng}
\email{p.hoyng@sron.nl}
\affiliation{SRON Netherlands Institute for Space Research, \\
Sorbonnelaan 2, 3584 CA Utrecht, The Netherlands}

\date{\today}


\begin{abstract}
We analyze the dynamics of a driven, damped pendulum as used in clocks. We  
derive equations for the amplitude $\lambda$ and the phase $\psi$ of the 
oscillation, on time scales longer than the pendulum period. The equations 
are first order ODEs and permit fast simulations of the joint effects of 
circular and escapement errors and friction for long times (1 year real time in 
a few minutes). The equations contain two averages of the driving torque over 
a period, so that the results are not very sensitive to the `fine structure' 
of the driving. We adopt a constant-torque escapement and study (1) the 
stationary pendulum rate $\dot\psi$ as a function of driving torque and 
friction, (2) the reaction of the pendulum to a sudden change in the driving 
torque, and (3) to stationary noisy driving. The equations for $\lambda$ and 
$\psi$ are shown to describe the pendulum dynamics quite well on time scales 
of a period and much longer. The emphasis is on a clear exposition of the 
physics.
\end{abstract}


\maketitle


\section{Introduction}
\label{sec:intro}
\noindent
Suppose you wish to study the long-term time keeping properties of that 
longcase clock that for years on end has been peacefully eating away the 
seconds of your life. How would you do that? Making regular measurements 
is one thing, and numerical simulations is another. In the latter case you 
are in for a surprise: you have to solve the equation of a driven, damped 
harmonic oscillator in steps much smaller than the period of the pendulum, 
allowing for the slow changes (in driving, friction, temperature,..). And 
although each next oscillation is virtually identical to the previous one, 
you must follow them all in great detail to get any accuracy in the final 
result. This leads to long simulations, even in these days of fast desktop 
computers. Here we develop and validate a much faster method, that allows 
making time steps of the size of the pendulum period. 

There is an extensive literature on the various types of pendulum and the 
effect of perturbations on their motion. Rawlings \cite{RAW80} and Woodward
\cite{W06} expound the scientific principles underlying clocks and pendulums, 
while Baker and Blackburn \cite{BB05} review the diversity of pendulums 
occurring in nature. A useful (but incomplete) compilation of the literature 
has been given by Gauld. \cite{CG04}

The oldest known pendulum timing error is the circular error, i.e. the fact 
that the period of a free pendulum depends weakly on its amplitude. Amplitude 
variations induce therefore small timing errors. Huygens showed in 1657 how 
these may be eliminated by the use of cycloidal cheeks. \cite{RAW80} 
Henceforth we shall often just speak of errors instead of timing errors. The 
escapement, \cite{RAW80} i.e. the mechanism that transfers energy to the 
pendulum to keep it swinging, is an unavoidable source of errors. These have 
been computed by Airy in 1830 in a seminal paper that preludes all later 
developments on this topic. \cite{GBA30} 

Kesteven \cite{MK78} computed the period change of a pendulum driven by a 
dead-beat or a gravity escapement, while Nelson and Olsson \cite{NO86} did so
for air drag and buoyancy. Denny \cite{MD02} shows how a simple escapement 
makes a pendulum settle in a stable limit cycle, and presents a useful summary  
of Airy's method. Friction is often modelled as a force proportional to the 
velocity, but Andronov {\em et al.}\cite{AVK66} treat several other types of 
friction.

Advanced models recognize the fact that a pendulum with its escapement is a 
mechanical system with several degrees of freedom that may be analyzed using 
so-called impulsive differential equations, see e.g. Moline {\em et al.} 
\cite{MWV12} and Moon and Stiefel. \cite{MS06} On the practical side, Matthys 
\cite{M07} gives detailed suggestions for the design and construction of a 
precision pendulum, with much attention to the properties and choice of 
materials.   

The basis for a study of a pendulum's time keeping properties is the equation 
of motion. It incorporates all effects but requires a time step much smaller 
than the period and contains therefore too much information. Here we derive 
equations for the pendulum dynamics on a meso-time scale, i.e. the time scale 
of a period. This work amounts to a new, more general formulation of 
escapement theory. It permits much faster simulations, while all timing error 
sources operate in concert.

The technique is explained in Secs.~\ref{sec:amph} and \ref{sec:prac}. We 
introduce a simple model escapement in Sec.~\ref{sec:modesc}, and analyse the
stationary operation of a pendulum driven by this escapement in 
Sec.~\ref{sec:statop}. In Secs.~\ref{sec:such} and \ref{sec:perst} we apply 
the theory to study the reaction of the pendulum to a sudden parameter change 
and to stationary noisy driving. The second novel aspect of our work is that 
we validate our approach in detail by comparing the results with data computed 
directly from the equation of motion. We discuss our results in 
Sec.~\ref{sec:disc}.


\section{equations for the amplitude and phase}
\label{sec:amph}
The pendulum oscillates in a plane, making an angle $\varphi(t)$ with the 
vertical. The equation of motion for $\varphi$ is  
\begin{equation}
\ddot\varphi+2\varepsilon\dot\varphi+\sin\varphi= m(\varphi)\ .
\label{eq:ddt1}
\end{equation}
Here $\varphi$ is measured in radians, and $\varepsilon$ and $m$ are the 
friction coefficient and the driving torque (divided by the moment of inertia    
of the pendulum). The dot stands for the time-derivative: $\dot{ }=\md/\md t$, 
$\ddot{ }=\md^2/\md t^2$, etc. Time has been scaled so that the period of 
unperturbed, small-amplitude oscillations is $2\pi$ ($\varepsilon=m=0$; 
$\sin\varphi\simeq\varphi$). We shall refer to this as the nominal period.


\begin{figure}
\centerline{\includegraphics[width=4.5cm]{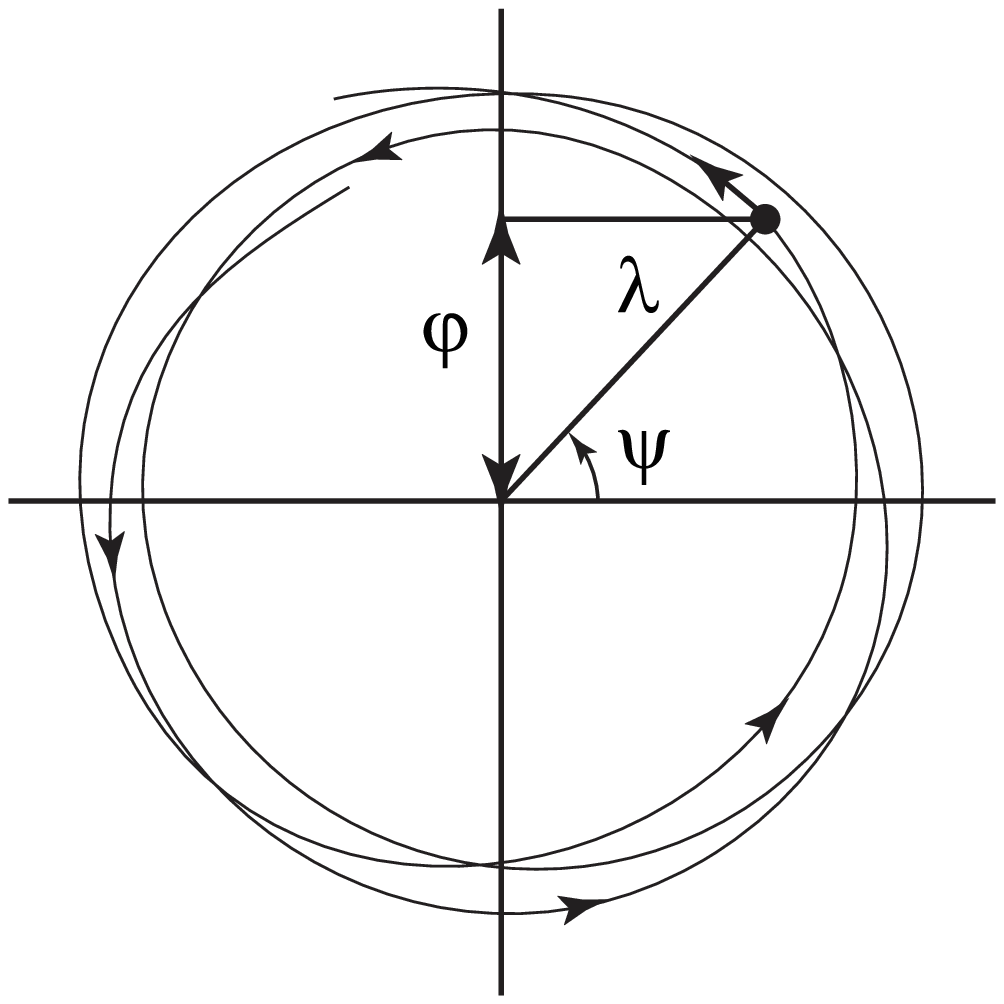}}
\caption{\small The motion of the pendulum is represented by a point with 
radius vector $\lambda$ rotating with angular speed $\dot\psi$, projected on 
the vertical axis, so that $\varphi=\lambda\sin\psi$. In this way we achieve 
that $\varphi=0$ when $\psi$ is an integer multiple of $\pi$. Both $\lambda$ 
and $\psi$ exhibit slow variations (slower than indicated), resulting in 
gradual changes in the swing amplitude and the period of the pendulum.}
\label{fig:pslaph}
\end{figure}


The escapement \cite{RAW80} is designed to deliver a torque $m$ that depends 
only on the angle $\varphi$. It has no intrinsic frequency of its own, 
but just delivers the torque required by the value of $\varphi$, irrespective 
of time. So, $m=m(\varphi)$ and Eq.~(\ref{eq:ddt1}) is autonomous. In actual 
fact $m$ may have a weak dependence on $\dot\varphi$ as well. We shall not 
consider the more general case $m=m(\varphi,\dot\varphi)$, and we shall take 
$\varepsilon$ constant. But the theory developed below can handle 
$\varepsilon$ and $m$ with arbitrary functional dependence on $\varphi$ and 
$\dot\varphi$.

It is straightforward to solve Eq.~(\ref{eq:ddt1}) numerically, but the result 
contains too much information for our purposes. The time keeping is determined 
by the times of successive $\varphi=0$ crossings, for which it is sufficient 
to know $\varphi(t)$ with a time resolution of about a period. In other words, 
we are interested in slow changes of the amplitude and phase. The problem of 
gradual secular changes in quasi-periodic systems has come up in physics time 
and again. One of the earliest applications was the problem of secular 
perturbations in planetary orbits. \cite{AM88} Here we shall follow the theory 
developed by Krylov, Bogoliubov and Mitropolski (KBM). \cite{BM61,BLV04} The 
idea is to take 
\begin{equation}
\varphi=\lambda\sin\psi\ ;\quad\dot\varphi=\lambda\cos\psi\ , 
\label{eq:deflabpsi}
\end{equation}
where $\lambda$ is the maximum of $\varphi$ (referred to as `the amplitude'), 
and $\psi$ the phase of the pendulum, see Fig.~\ref{fig:pslaph}. We may regard 
$\psi$ as the time measured by the pendulum (`pendulum time'), while $t$ is a 
stable reference time (`laboratory time'). One might think that the first 
relation $\varphi=\lambda\sin\psi$ suffices, for we may compute $\dot\varphi= \dot\lambda\sin\psi+\lambda\dot\psi\cos\psi$ and then $\ddot\varphi=\cdots$, 
insert all that in Eq.~(\ref{eq:ddt1}) and make approximations using $\lambda, 
\,\varepsilon\ll 1$. However, that path is fraught with ambiguities due to the 
fact that $\varphi=\lambda\sin\psi$ does not define $\lambda$ and $\psi$ 
uniquely. We do need a second relation.


\subsection{The method of Krylov, Bogoliubov and Mitropolski (KBM)}
\label{sec:KBR}
The next step is to solve Eq.~(\ref{eq:deflabpsi}) for $\lambda$ and $\psi$,
\begin{equation}
\lambda=(\varphi^2+\dot\varphi^2)^{1/2}\ ;\quad 
\psi=\arctan(\varphi/\dot\varphi)\ ,
\label{eq:laps}
\end{equation}
and to compute the derivatives:
\begin{eqnarray}
\dot\lambda\!&=&\!\frac{\varphi\dot\varphi+\dot\varphi\ddot\varphi}
{(\varphi^2+\dot\varphi^2)^{1/2}}\,=\,\frac{1}{\lambda}\,
(\ddot\varphi+\varphi)\dot\varphi\ ,
\label{eq:dotl1}  \\[3.mm]
\dot\psi\!&=&\!\frac{\dot\varphi^2-\varphi\ddot\varphi}
{\varphi^2+\dot\varphi^2}\ \ \ \ \ 
=\,1-\frac{1}{\lambda^2}\,(\ddot\varphi+\varphi)\varphi\ .
\quad
\label{eq:dotpsi1}
\end{eqnarray}
The derivative $\dot\psi$ of the phase is the rate at which the pendulum 
ticks. And $\hlf\lambda^2$ is the total (kinetic + potential) energy of the 
pendulum. Both right hand sides contain $\ddot\varphi+\varphi$, which the 
equation of motion (\ref{eq:ddt1}) allows us to write as: 
\begin{equation}
\ddot\varphi+\varphi=-2\varepsilon\dot\varphi+
(\varphi-\sin\varphi)+m(\varphi)\ .
\label{eq:ddt2} 
\end{equation}
On the right we recognize the terms responsible for the effects of friction, 
circular error and driving, respectively. We insert Eq.~(\ref{eq:ddt2}) in 
Eqs.~(\ref{eq:dotl1}) and (\ref{eq:dotpsi1}), and eliminate the remaining 
$\varphi$ and $\dot\varphi$ with the help of Eq.~(\ref{eq:deflabpsi}):
\begin{eqnarray}
\dot\lambda\!&=&\!-2\varepsilon\lambda\cos^2\psi+\{\lambda\sin\psi- 
\sin(\lambda\sin\psi)\}\,\cos\psi 
\nonumber \\[1.mm]
&&\qquad\qquad\qquad\qquad\qquad+\,m\cos\psi\,,
\label{eq:dotl2}  \\[3.mm]
\dot\psi-1\!&=&\!2\varepsilon\cos\psi\sin\psi+\{\lambda^{-1}\sin(\lambda
\sin\psi)-\sin\psi\}\,\sin\psi 
\nonumber \\[2.mm]
&&\qquad\qquad\qquad\qquad\qquad-\,(m/\lambda)\,\sin\psi\ .
\label{eq:dotpsi2}
\end{eqnarray}
Equations~(\ref{eq:dotl2}) and (\ref{eq:dotpsi2}) are still equivalent to 
Eq.~(\ref{eq:ddt1}). KBM then average over one period to remove the fast time 
scales. This amounts to integrating the equations over $\psi$ over the 
interval $[\psi_0-\pi,\,\psi_0+\pi]$ centered at an arbitrary $\psi_0$, and 
to assume that $\varepsilon$, $\lambda$ and $\dot\psi$ but not $m$, are 
constant over one period. Afterwards, the index 0 on $\psi$ is dropped. In 
doing so, the terms $\{\lambda\sin\psi-\sin(\lambda\sin\psi)\}\,\cos\psi$ in 
Eq.~(\ref{eq:dotl2}) and $2\varepsilon\cos\psi\sin\psi$ in Eq.~(\ref{eq:dotpsi2}) 
vanish because they are antisymmetric in $\psi$. Since $(2\pi)^{-1} 
\smallint_{-\pi} ^{\pi}\cos^2\psi\,\md\psi=\hlf$ we obtain:
\begin{eqnarray}
\dot\lambda&=&-\varepsilon\lambda+\oint\,m(\varphi)\cos\psi\,\md\psi\,+\,
O_1(\varepsilon^2)\ ,
\label{eq:dotl3}  \\[2.mm]
\dot\psi-1&=&-\frac{1}{16}\,\lambda^2-\frac{1}{\lambda}\, 
\oint\,m(\varphi)\sin\psi\,\md\psi\,+\,O_2(\varepsilon^2)\ ,
\quad
\label{eq:dotpsi3}
\end{eqnarray}
where we use the notation $\oint=(2\pi)^{-1}\smallint_{-\pi}^{\pi}\,$. We 
demonstrate in appendix~\ref{sec:circer} how the circular error term 
$-\lambda^2/16$ is extracted from Eq.~(\ref{eq:dotpsi2}). 

In reality $\varepsilon$, $\lambda$ and $\dot\psi$ are not constant over a 
period. Bogoliubov and Mitropolski \cite{BM61} show that this produces 
correction terms in Eqs.~(\ref{eq:dotl3}) and (\ref{eq:dotpsi3}) having the 
form of a power series in $\varepsilon$. These higher order terms play no role 
here, except in Sec.~\ref{sec:perst} where we believe to observe the effect 
of the lowest order terms $\propto\varepsilon^2$.   

Briefly, what if $\varepsilon$ and $m$ depend on $\varphi$ and $\dot
\varphi$? Take for example $\varepsilon$. Then set $\varepsilon=\varepsilon (\varphi,\dot\varphi)=\varepsilon(\lambda\sin\psi,\lambda\cos\varphi)$ in 
Eqs.~(\ref{eq:dotl2}) and (\ref{eq:dotpsi2}). The average over a period will now 
generate a friction-like term in the rate equation (\ref{eq:dotpsi3}), etc.


\section{The physics of eqs.~(\ref{eq:dotl3}) and (\ref{eq:dotpsi3})}
\label{sec:prac}
Equations (\ref{eq:dotl3}) and (\ref{eq:dotpsi3}) are first order ODEs 
well-suited for a study of the long-term dynamics of the pendulum. 
Equation (\ref{eq:dotl3}) determines the amplitude $\lambda$ and the total energy 
$\lambda^2/2$ of the pendulum. The amplitude changes only slowly because the 
friction time scale $\varepsilon^{-1}$ is long, $1/(2\pi\varepsilon)$ nominal 
periods, and because the escapement needs to deliver many small kicks to alter 
the energy $\hlf\lambda^2$. The damping is compensated by the driving term 
$\oint m\cos\psi\,\md\psi$. A back-of-the-envelope derivation of 
Eqs.~(\ref{eq:dotl3}) and (\ref{eq:dotpsi3}) is given in 
appendix~\ref{sec:appeqs}.


\subsection{The rate equation (\ref{eq:dotpsi3})}
\label{sec:rateq}
Equation~(\ref{eq:dotpsi3}) for the rate features the {\em circular error} term 
$-\lambda^2/16$, due to the difference between $\varphi$ and $\sin\varphi$ 
in Eq.~(\ref{eq:ddt2}). It has a minus sign because the circular error 
increases the period, so it must reduce the rate. Friction does not appear, 
and has therefore only an indirect effect on the rate through its influence 
on $\lambda$. 

The term $-\lambda^{-1}\oint m\sin\psi\,\md\psi$ is called the {\em escapement 
error}. It specifies the relative rate change induced by a driving torque $m$,  
as computed already by Airy \cite{GBA30}. The history of horology is to a 
large extent a quest for technical solutions to make this term and the 
circular error as small as possible. 

An important feature of both equations is that the driving torque $m$ appears 
only in the form of two averages over a period. This is a consequence of the 
elimination of fast time scales and demonstrates that much of the fine 
structure in the driving torque averages out and is irrelevant for the 
pendulum's long term dynamics. Only two numbers matter, namely the even 
(cosine) and odd (sine) moment, and all pendulums with escapements having 
the same $\oint m\cos\psi\,\md\psi$ and $\oint m\sin\psi\,\md\psi$ are in 
principle equivalent time keepers. 


\begin{figure}
\vspace{-2.mm}
\centerline{\includegraphics[width=9.cm]{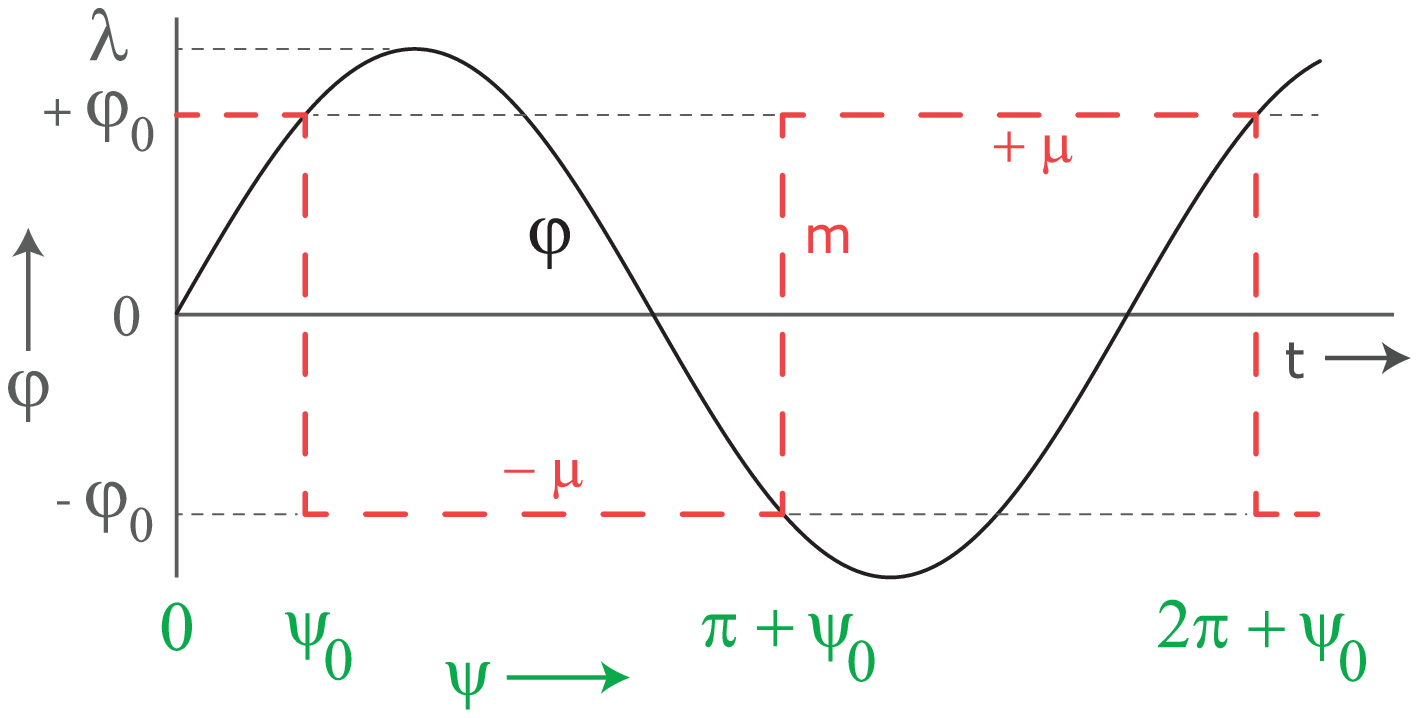}}
\vspace{-5.mm}
\caption{\small The model escapement delivers a constant torque $m=\mu$, 
switching to $m=-\mu$ after half a period ($-\;-\;-$). The switch is at a 
fixed pendulum angle $\varphi_0$ in the first quarter of the swing, and 
$-\varphi_0$ in the third. The phase $\psi$ is indicated at the bottom.} 
\label{fig:modesc}
\end{figure}


\subsection{A model escapement}
\label{sec:modesc}
For our numerical work we introduce a simple escapement, see 
Fig.~\ref{fig:modesc} and Woodward \cite{W06chap2}. A constant driving 
torque $m=\mu>0$ switching to $m=-\mu$ after half a period, and so on. 
The switch is at a fixed (mechanically determined) pendulum angle $\varphi_0$. 
The phase $\psi_0$ at that moment depends on the amplitude according to 
Eq.~(\ref{eq:deflabpsi}):
\begin{equation}
\varphi_0=\lambda\sin\psi_0\ ;\qquad 0<\psi_0<\pi/2\ .
\label{eq:defdel}
\end{equation}
The swing amplitude $\lambda$ must be larger than $\varphi_0$, otherwise the 
driving torque cannot switch sign, and the pendulum will stall. Since 
$\varphi_0=\lambda\sin\psi_0$ has two solutions, $\psi_0$ and $\pi-\psi_0$, we 
impose $0<\psi_0<\pi/2$ because escapements generally switch to a torque 
counteracting the motion before the pendulum reaches maximum amplitude.

Actual escapements deliver a more complicated $m$, \cite{MK78,MS06,MWV12} but 
as argued above, this fine structure of $m$ is expected to be relatively 
unimportant for the performance. So we hope to catch the main features of this 
and similar escapements with the help of this simple model. We compute the two 
averages required in Eqs.~(\ref{eq:dotl3}) and (\ref{eq:dotpsi3}):
\begin{eqnarray}
a\!&\equiv&\!\!\oint m\,\cos\psi\,\md\psi   
\label{eq:ca1} \\[1.mm]
&=&\!\!\frac{1}{2\pi}\,\bigg\{-\int_{\psi_0}^{\pi+\psi_0}
    +\int_{\pi+\psi_0}^{2\pi+\psi_0}\bigg\}\,\mu\cos\psi\,\md\psi  
\nonumber  \\[4.mm]
&=&\!\!\frac{2\mu}{\pi}\,\sin\psi_0\,=\,\frac{2\mu\varphi_0}{\pi\lambda}\ .
\label{eq:ca2}
\end{eqnarray}
Since the integrand is a periodic function of $\psi$ we may start the  
integration where we like, as long as we integrate over a full period. A 
similar computation yields 
\begin{equation}
b\equiv\oint m\,\sin\psi\,\md\psi=-\,\frac{2\mu}{\pi}\,\cos\psi_0\ . 
\qquad\ \ 
\label{eq:cb1}
\end{equation}
Equations (\ref{eq:dotl3}) and (\ref{eq:dotpsi3}) for the amplitude and the 
phase may now be written as:
\begin{eqnarray}
\frac{\md\lambda^2}{\md t}&=&-2\varepsilon\lambda^2+\frac{4\mu\varphi_0}{\pi}
\,+\,O_1(\varepsilon^2)\ ,
\label{eq:dotl4}  \\[2.mm]
\dot\psi-1&=&-\frac{1}{16}\,\lambda^2+\frac{2\mu}{\pi\lambda}\,\cos\psi_0
\,+\,O_2(\varepsilon^2)\ ,
\quad
\label{eq:dotpsi4}
\end{eqnarray}
where Eq.~(\ref{eq:dotpsi4}) may be further reduced with $\cos\psi_0=[1-
(\varphi_0/\lambda)^2]^{1/2}$, as follows from Eq.~(\ref{eq:defdel}). Due to 
a different notation the escapement error does not look like any of the 
expressions in Ch.~7 of Rawlings \cite{RAW80}. But we have verified that we 
can reproduce these formulae starting from our expressions $[2\mu/(\pi
\lambda)]\,\cos\psi_0$ in Eq.~(\ref{eq:dotpsi4}) or $-(1/\lambda)\oint 
m\sin\psi\,\md\psi$ in Eq.~(\ref{eq:dotpsi3}). So the difference is only 
apparent. In appendix~\ref{sec:appblow} we analyse the action of an escapement 
that delivers a single impulse in each period. 


\subsection{Analytic solution}
\label{sec:ansol}
Before we start using Eqs.~(\ref{eq:dotl4}) and (\ref{eq:dotpsi4}), we point 
out that they often possess a simple analytic solution, which may be useful 
in theoretical investigations. For example, if $\varepsilon$ is constant and 
the transient responses are over, the system is in a quasi-stationary state 
and the solution of Eq.~(\ref{eq:dotl4}) is: 
\begin{equation}
\lambda^2=\frac{4\varphi_0}{\pi}\int_0^\infty\exp(-2\varepsilon\tau)\,
\mu(t-\tau)\,\md\tau\ .
\label{eq:lamanal}
\end{equation}
This permits us to compute $\lambda(t)$ for any time-dependent driving torque 
$\mu(t)$. And once $\lambda$ is known, we may compute $\dot\psi$ from 
Eq.~(\ref{eq:dotpsi4}). For example, let there be a jump in $\mu$ at $t=0$ and 
$\mu=\mu_0$ for $t<0$ and $\mu=\mu_1$ for $t>0$. Inserting that in 
Eq.~(\ref{eq:lamanal}) produces
\begin{equation}
\displaystyle\lambda^2\,=\,\frac{2\varphi_0}{\pi\varepsilon}\cdot\left\{\!
\begin{array}{ll} 
\mu_0 & \ t<0\ , \\[3.mm]
\{\mu_1-(\mu_1-\mu_0)\exp(-2\varepsilon t)\} & \ t>0\ .  
\end{array}
\right.
\label{eq:l2jump}
\end{equation}
This result will be used later to explain the numerical work in 
Sec.~\ref{sec:such}. Just replace $t$ by $t-t_0$ everywhere in 
Eq.~(\ref{eq:l2jump}) if the jump is at $t_0$.


\begin{figure}
\vspace{-3.mm}
\centerline{\includegraphics[width=9.cm]{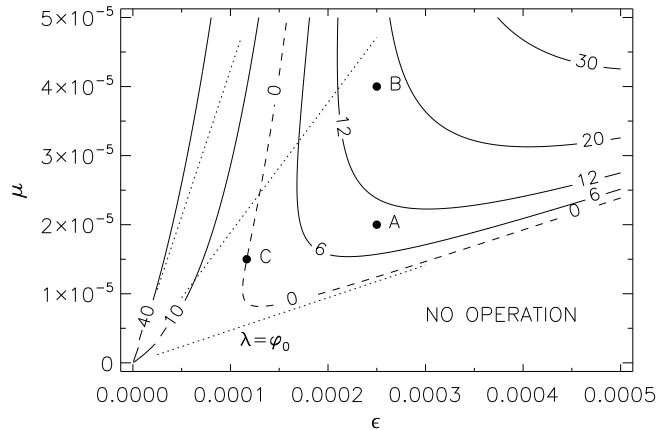}}
\vspace{-3.mm}
\caption{\small Contour plot of $\dot\psi-1$ in sec/day as a function of 
$\varepsilon$ and $\mu$, for $\varphi_0=0.03\cong 1.7^\circ$. This is the 
rate in excess of the nominal rate of a pendulum driven by the model 
escapement. Lines of constant swing amplitude are dotted, $\lambda=\varphi_0$ 
and higher up $2\varphi_0$ and $3\varphi_0$. For example, if $\varepsilon\sim 
5\times 10^{-5}$ and $\mu\sim 10^{-5}$, the pendulum loses about $10\,$s per 
day with respect to the nominal rate, and the swing amplitude is about 
$2\varphi_0$. Further details in main text.}
\label{fig:rate}
\end{figure}


\section{Stationary operation}
\label{sec:statop}
The asymptotic theory developed above enables us to address many questions 
that would be difficult to answer if we had only the equation of motion 
(\ref{eq:ddt1}). We begin with the case that $\varepsilon$, $\mu$ and 
$\varphi_0$ are constant. The pendulum will then perform stationary 
oscillations after some time. We study how the stationary amplitude $\lambda$ 
and rate $\dot\psi$ depend on the parameters. The amplitude follows by setting 
$\md\lambda^2/\md t=0$ in Eq.~(\ref{eq:dotl4}):
\begin{equation}
\lambda=\varphi_0\bigg(\frac{2\mu}{\pi\varepsilon\varphi_0}\bigg)^{1/2}\ ,
\label{eq:statam}
\end{equation}
and using this in Eq.~(\ref{eq:dotpsi4}) we obtain the stationary rate:
\begin{equation}
\dot\psi-1\,=\,-\,\frac{\mu\varphi_0}{8\pi\varepsilon}\,+\,\varepsilon
\bigg(\frac{2\mu}{\pi\varepsilon\varphi_0}-1\bigg)^{1/2}.
\label{eq:statrate}
\end{equation}  
The first term on the right is the circular error (no longer readily 
recognizable as such). It is negative and reduces the rate. The second term 
is the escapement error, which is manifestly positive. It follows that the 
driving necessary to keep the pendulum going, always forces it to run a little 
faster as well. This confirms one's intuition, and the physical reason is, 
loosely speaking, that during most of the time the escapement is pushing the 
pendulum ever so gently forward. These statements hold for the model 
escapement and comparable types, e.g. the recoil escapement, \cite{W06recoil} 
but not for all, see appendix~\ref{sec:appblow}.

Fig.~\ref{fig:rate} shows the excess rate on top of the nominal rate $\dot\psi
=1$ in sec/day due to circular and driving error together. The 
$\varepsilon$-$\mu$ plane is divided in three regions: \\
1. A losing rate, to the left and up. In this region the amplitude $\lambda$ 
is large because the friction is relatively small. As a result, the circular 
error is larger than the escapement error, and the pendulum rate is 
sub-nominal. \\
2. A gaining rate, to the right and up: here the effect of the driving torque 
dominates (escapement error). These regions are separated by the dashed 
contour where circular and escapement error compensate each other. Pendulums 
operating on this dashed line run at nominal rate. \\ 
3. In the third region, below the line $\lambda=\varphi_0$, the driving is 
too weak to keep the swing amplitude $\lambda$ above the value $\varphi_0$, 
and the pendulum stalls. The minimum required torque $\mu$ is $\hlf\pi
\varepsilon\varphi_0$.


\subsection{Losing or gaining?}
\label{sec:losgain}
Here comes a perennial question: will a clock lose or gain when a parameter 
is changed? Take for example the driving torque. The discussion usually 
proceeds as follows: $\mu\uparrow$ implies a larger amplitude $\lambda$, which 
means a larger circular error, i.e. $\dot\psi\downarrow$. But sometimes we 
observe the opposite. How come? On closer look, the answer depends on the 
pendulum's initial operating point in Fig.~\ref{fig:rate}. If that is towards 
the left where the circular error dominates, then $\mu\uparrow$ makes the 
first term (circular error) in Eq. (\ref{eq:statrate}) go down and 
$\dot\psi\downarrow$. But if we start more to the right where the escapement 
error dominates, the second term in Eq. (\ref{eq:statrate}) (escapement error) 
increases. So $\dot\psi\downarrow$ when we start on the left in 
Fig.~\ref{fig:rate}, but $\dot\psi\uparrow$ when we begin on the right. There 
is no simple answer.

Variations in friction may be dealt with in the same manner. A remarkable 
feature is that the pendulum may run faster when friction is increased. Take 
for example $\mu$ fixed at $2\times 10^{-5}$ in Fig.~\ref{fig:rate}, then the 
rate goes up with $\varepsilon$ until $\varepsilon\sim 2.5\times 10^{-4}$ and 
decreases only thereafter. 
 

\subsection{Accuracy}
\label{sec:oomest}
The accuracy of the time keeping is a subject with many different aspects, of 
which we shall discuss only one. Experience has shown that the accuracy of a 
pendulum is largely set by the friction coefficient $\varepsilon$. The driving 
plays a minor role. Why is that so? The answer comes in two steps.

We focus on long time scales ($\gg\varepsilon^{-1}$). In that case we may 
ignore the time derivative in Eq.~(\ref{eq:dotl3}) and obtain the following 
order-of-magnitude estimate: $\varepsilon\lambda\sim m$, or $m/\lambda\sim
\varepsilon$, which we use to estimate $\dot\psi-1$ with 
Eq.~(\ref{eq:dotpsi3}):
\begin{equation}
\dot\psi-1\,\sim\,m/\lambda\,\sim\,\varepsilon\ .
\label{eq:ratest}
\end{equation}
Here we assume that the pendulum operates in region 2 of Fig.~\ref{fig:rate} 
where the circular error is negligible. Thus we have found that $\dot\psi-1
\sim\varepsilon$, and Eq.~(\ref{eq:statrate}) is basically telling the same 
story. The point is that the driving torque no longer appears in 
Eq.~(\ref{eq:ratest}) because it could be approximately eliminated by requiring 
quasi-stationary operation. 

The second part of the answer is that as long as the parameters are constant, 
nothing changes. The rate is exactly constant and the pendulum is a perfect 
time keeper, even though it may not run at nominal rate. But parameters are 
never constant and change on a variety of time scales. Experience has shown 
that changes of several tens of percent in the parameters and more are not 
uncommon. And that makes $\varepsilon$ a reasonable upper limit for the 
relative precision of the pendulum rate.


\subsection{Optimal driving torque}
\label{sec:optdriv}
From Eqs.~(\ref{eq:dotl3}) and (\ref{eq:dotpsi3}) the least disturbing driving 
torque is seen to be an impulse in a narrow phase interval $\delta\psi\ll 1$ 
centered at $\psi=k\pi$, when the pendulum passes through its equilibrium 
position $\varphi=0$. The escapement error is then $\lambda^{-1}\oint m\sin
\psi\,\md\psi\sim(m/\lambda)(\delta\psi)^2\sim\varepsilon\delta\psi$. Here we 
used that $m\delta\psi\sim\varepsilon\lambda$, as follows from 
Eq.~(\ref{eq:dotl3}). We conclude that it should be possible to beat the 
precision estimate (\ref{eq:ratest}) by a factor $\delta\psi\ll 1$, with the 
help of a specially designed escapement. All this is well known - we merely 
illustrate that the present formulation which uses concepts that are close to 
the instrument, allows a more direct answer to such questions than 
Eq.~(\ref{eq:ddt1}) alone. 


\begin{figure}
\vspace{-25.mm}
\centerline{\includegraphics[width=8.cm]{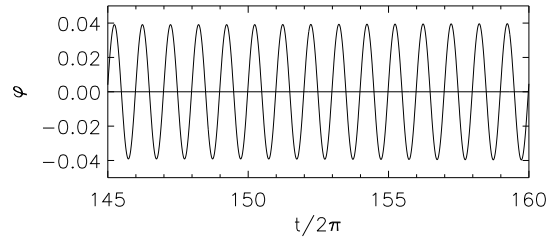}}
\caption{\small Numerically determined pendulum angle $\varphi$. On the 
horizontal axis time in nominal periods. Parameters: $\varepsilon=2.5\times 
10^{-4}$, $\varphi_0=0.03$, timestep $dt=0.03$. Initial conditions: 
$\varphi(0)=0$, $\dot\varphi(0)=(2\mu_0\varphi_0/\pi\varepsilon)^{1/2}$ 
(stationary swing). At $t/2\pi=150$ the driving torque switches from 
$\mu_0=2\times 10^{-5}$ to $\mu_1=4\times 10^{-5}$. There is no visible 
reaction.}
\label{fig:jump_zoom}
\end{figure}


\section{Transient effects}
\label{sec:such}
We validate the asymptotic theory developed above with two case studies. 
The first is the reaction of the pendulum to a jump in the driving torque. 
We solve Eq.~(\ref{eq:ddt1}) numerically, parameters as in 
Fig.~\ref{fig:jump_zoom}; for details see appendix~\ref{sec:appnum}.
The pendulum is set off in a stationary swing, and the value of $\dot
\varphi(0)$ follows from Eq.~(\ref{eq:laps}): $\dot\varphi(0)=\lambda$ and 
$\lambda$ may be had from Eq.~(\ref{eq:statam}). There are $2\pi/dt\sim 209$ 
time steps in one nominal period. A similar numerical experiment has been 
carried out by Woodward. \cite{W06sim}

The solution, Fig.~\ref{fig:jump_zoom}, is not very telling: a series of 
identical oscillations, and there is no visible reaction in the vicinity of 
the jump in $\mu$ at $t/2\pi=150$. To refine the analysis we determine the 
times $t_i$ that $\varphi(t)$ crosses zero (from $-$ to $+$). Numerical 
details are given in appendix~\ref{sec:appnum}. Since $\psi$ increases by 
$2\pi$ at each next zero $i$ of $\varphi$ (by definition), we may determine 
the time (i.e. phase) $\psi$ and the rate $\dot\psi$ measured by the pendulum 
on a non-equidistant grid $\{t_i\}$, as follows: 
\begin{equation}
\psi_i=2\pi i\,;\qquad\dot\psi_i=2\pi(t_i-t_{i-1})^{-1}\ .
\label{eq:ratemeas}
\end{equation}
This procedure can be easily implemented in the numerical model (and in a 
pendulum experiment). The measured rates may then be confronted with theory, 
Eq.~(\ref{eq:dotpsi4}). 

\begin{figure}
\vspace{-7.mm}
\centerline{\includegraphics[width=13.cm]{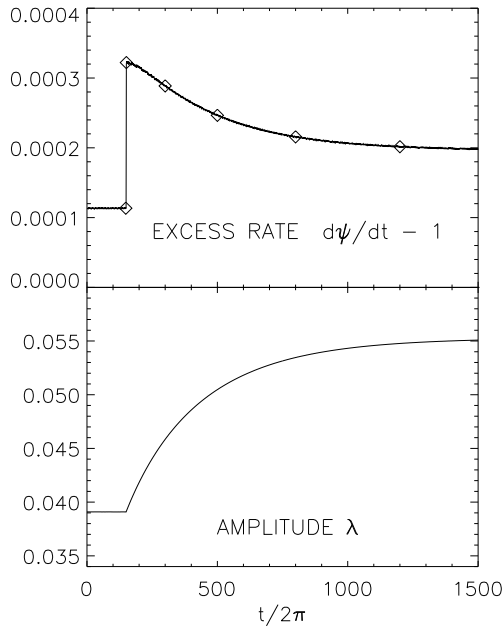}}
\caption{\small Reaction of the pendulum to a jump in the driving torque 
(parameters as in Fig.~\ref{fig:jump_zoom}). On the horizontal axis time in 
nominal periods. At $t/2\pi=150$ the driving torque switches from $\mu_0=
2\times 10^{-5}$ to $\mu_1=4\times 10^{-5}$. Top panel: excess rate of the 
pendulum (in seconds per sec), as `observed', that is, computed from 
definition (\ref{eq:ratemeas}). The noise on the rate curve is numerical and 
can be reduced by taking a smaller timestep $dt$. Diamonds $\diamond$ are 
values inferred from theory, Eq.~(\ref{eq:dotpsi4}). Bottom panel: amplitude 
$\lambda$ computed from Eq.~(\ref{eq:laps}) and the simulation data; 
Eq.~(\ref{eq:l2jump}) yields the same result.}
\label{fig:jump}
\end{figure}

The reaction of the pendulum, Fig.~\ref{fig:jump}, top panel, is now clearly 
visible. The initial operating point of the pendulum is $A$ in 
Fig.~\ref{fig:rate}, where it gains about $1.1\times 10^{-4}$ or $\sim 10$ 
sec/day. The rate changes instantaneously, together with the jump in the 
driving, and then tapers off to a new stationary value $\sim 2\times 10^{-4}$ 
or $\sim 17$ s/day in the new operating point $B$. The time scale is set by 
$\lambda$, Fig.~\ref{fig:jump}, bottom panel, and $\lambda$ changes only 
slowly because it takes many small pushes of the escapement to change the 
energy $\lambda^2/2$. Once $\lambda$ is known, we compute $\dot\psi-1$ with 
Eq.~(\ref{eq:dotpsi4}), as a verification. A few values are displayed as 
$\diamond$ in Fig.~\ref{fig:jump}. 

The magnitude of the rate jump follows directly from Eq.~(\ref{eq:dotpsi4}): since 
$\lambda$ changes slowly, and $\cos\psi_0$ depends only on $\lambda$, the rate 
difference just before and after the jump is $\delta\dot\psi=2(\delta\mu)\cos
\psi_0/(\pi\lambda)$. This allows us to compute that $\dot\psi-1=3.2\times 
10^{-4}$ after the jump, as observed. The jump in the rate is tiny but well 
measurable. It is an arresting experience for students experimenting with a 
pendulum in a practical physics course. The unaided eye sees no change when 
the driving is altered, and yet it happens, just because the escapement pushes 
a little harder.


Finally, we compute the time shift the pendulum incurs. Locating the jump for 
simplicity at $t=0$, the phase shift is $\Delta\psi=\int_0^\infty(\dot\psi-
\dot\psi_{\rm r})\,\md t$, where $\dot\psi$ is the rate given by 
Eq.~(\ref{eq:dotpsi4}) using Eq.~(\ref{eq:l2jump}) for $\lambda$, and 
$\dot\psi_{\rm r}$ the constant rate of a reference pendulum operating in point 
$B$, i.e. the area under the curve $\dot\psi$ and above the tangent to 
$\dot\psi$ in $t=\infty$. We find $\Delta\psi=0.295$, or $(\Delta\psi/2\pi)
\cdot 2\,$s $\sim 0.1\,$s for a $1\,$m pendulum.

We have also looked at the pendulum's reaction to sudden changes in 
$\varepsilon$. The main difference is that the rate no longer makes a sudden 
jerk but adapts slowly to its final value, like $\lambda$ does. In conclusion 
we can say that the asymptotic theory seems to work well, also for a rapid 
change as we encountered here. 


\section{Long-term behavior}
\label{sec:perst}
The second case study is a pendulum with a variable driving torque, running 
at nominal rate in point $C$ in Fig.~\ref{fig:rate} in the absence of 
variability. The co-ordinates of $C$ are $\mu_0=1.5\times 10^{-5}$ and 
$\varepsilon=1.167245\times 10^{-4}$ (obtained by solving 
Eq.~(\ref{eq:dotpsi4}) for $\varepsilon$ with $\dot\psi=1$ and $\mu_0=1.5
\times 10^{-5}$). 

The reason for selecting this model is not because we think it is particularly 
realistic, but because we need a `rough' test bed to see how well the data 
generated by the `numerical clock', i.e. Eqs.~(\ref{eq:ddt1}) and 
(\ref{eq:dmu}), are recovered by the theoretical model, i.e. 
Eqs.~(\ref{eq:dotl4}) and (\ref{eq:dotpsi4}) together with (\ref{eq:dmu}). 

The simulation advances Eq.~(\ref{eq:ddt1}) for $10^9$ timesteps ($dt=0.03$), 
comprising $\sim 4.8\times 10^6$ nominal periods, or about 110 days for a 
$1\,$m pendulum. The initial conditions are as in Fig.~\ref{fig:jump_zoom} 
(stationary swing). During the simulation we determine the $-+$ zero crossing 
times $\{t_i\}$ of $\varphi$ and the pendulum time $\psi_{{\rm p},i}$ at these 
moments, basically by counting sine waves, see definition (\ref{eq:ratemeas}). 
We also integrate Eqs.~(\ref{eq:dotl4}) and (\ref{eq:dotpsi4}) on this 
$\{t_i\}$ grid, to determine the model pendulum time $\psi_{{\rm m},i}$. 

The driving torque $\mu$ has a random component: 
\begin{equation}
\mu=\mu_0\,\{1+\delta\mu(t)/\mu_0\}\ ,
\label{eq:dmu}
\end{equation}
where $\delta\mu(t)$ is a normally-distributed random variable with zero mean 
and variance $\delta\mu_\rms^2$ that is renewed at every $-+$ zero crossing of 
$\varphi$. The coherence or correlation time of $\delta\mu(t)$ is therefore one 
nominal period.

\begin{figure}
\vspace{2.mm}
\includegraphics[width=8.5 cm]{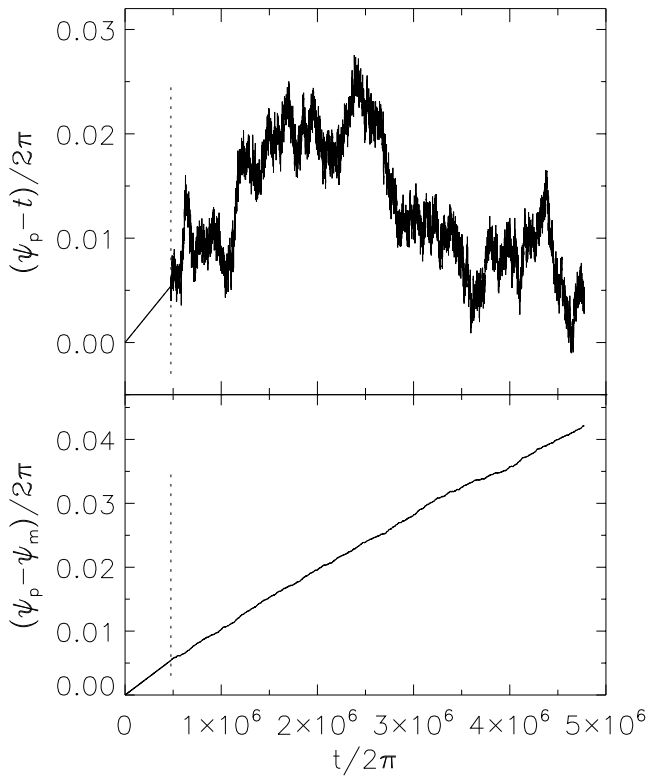}
\caption{\small Simulation of the pendulum motion comprising $10^9$ timesteps, 
or $10^9\,dt/2\pi\sim 4.8\times 10^6$ nominal periods, about $110$ days for a 
$1\,$m pendulum. Shown is the difference between pendulum time $\psi_{\rm p}$ 
and laboratory time $t$ (top), and between pendulum time $\psi_{\rm p}$ and 
model pendulum time $\psi_{\rm m}$ (bottom), as a function of laboratory time. 
To reduce blackening only one out of every 500 data points is plotted. 
Parameters: $\mu_0=1.5\times 10^{-5},\ \varepsilon=1.167245\times 10^{-4},\ 
\varphi_0=0.03,\allowbreak\ dt=0.03,\ \delta\mu_\rms=0.3\,\mu_0$. The random 
component $\delta\mu(t)$ in the driving is switched on beyond the dotted line 
($t/2\pi>4.8\times 10^5$). The simulation took $10^4\,$s (double precision 
IDL).}
\label{fig:quasistat}
\end{figure}

In spite of the location of $C$ in Fig.~\ref{fig:rate}, the pendulum does not 
run at nominal rate. In the absence of variability it gains $\sim 0.005/5
\times 10^{-5}=10^{-8}$, or $\sim 0.3\,$s/yr, see Fig.~\ref{fig:quasistat}, 
top panel. This is ten times better than the performance of a Shortt clock, 
see Ref.~\onlinecite{RAW80}, p. 126.
The effect is therefore extremely small, but we discuss it because we are 
after long-term effects here. Otherwise, the pendulum is seen to be almost 
impervious to the large variability in the driving, with an \rms\ gain of 
$\sim 0.04/4\times 10^6\sim 10^{-8}$, or $\sim 0.3\,$s/yr (same number by 
coincidence). 

Why is it so insensitive to the fluctuations? The decisive factor is the fact 
that the variations are fast, $\delta\mu(t)$ being renewed after each period, 
while the pendulum has a long memory $\sim\varepsilon^{-1}$, i.e. many 
periods. So there is a considerable cancellation of the random component of 
$\mu$. 

Slow variability however (time scale much larger than the period), results in 
little cancellation, so that the pendulum is quite sensitive to this type of 
variability. Basically, we rediscover here what horologists have known for a 
long time: accurate time keeping is all about controlling and eliminating 
slow, low-frequency variations in the parameters. Fast variations are much 
less detrimental.


Returning to the main question, how well do Eqs.~(\ref{eq:dotl4}) and 
(\ref{eq:dotpsi4}) model the pendulum dynamics? The bottom panel of 
Fig.~\ref{fig:quasistat} shows that the strong variability of pendulum time 
$\psi_{\rm p}$ seen in the top panel and induced by the driving noise, is by 
and large correctly modeled in $\psi_{\rm m}$ and removed. Only this small 
secular difference of $\sim 10^{-8}$ remains. We ran other simulations, and 
all graphs of $\psi_{\rm p}-\psi_{\rm m}$ versus $t$ looked more or less the 
same. This suggests that we see the effect of the $O_2(\varepsilon^2)$ term in 
Eq.~(\ref{eq:dotpsi4}), a term that we cannot model out as its theoretical 
form is unknown.\cite{BM61,BLV04} Perhaps we see the second-order correction 
$-\hlf\varepsilon^2$ of the oscillator frequency. \cite{LL78} That has the 
right order of magnitude and sign since $-\hlf\varepsilon^2\sim -10^{-8}$. 

Our conclusion is that also in this case the asymptotic theory appears to work 
well. Equations (\ref{eq:dotl4}) and (\ref{eq:dotpsi4}) deliver a relative rate 
precision of $\sim\varepsilon$ or better. This suffices for a physical 
pendulum as that attains a precision of $\sim\varepsilon$ anyway 
(Sec.~\ref{sec:oomest}). In numerical experiments where one may switch on and 
off perturbations at liberty, there may be cases that require computation and 
inclusion of the second order terms $O_1$ and $O_2$.


\section{Concluding remarks}
\label{sec:disc}
The purpose of this study is to demonstrate that the long-term dynamics of a 
pendulum as typically used in clocks can be simulated efficiently with two 
first order ODEs, Eq.~(\ref{eq:dotl4}) for the amplitude $\lambda$ and 
Eq.~(\ref{eq:dotpsi4}) for the phase $\psi$ of the pendulum. By using the rate 
$\dot\psi$ instead of its inverse (the period) and by averaging over fast time 
scales, we obtain simple model equations (\ref{eq:dotl4}) and 
(\ref{eq:dotpsi4}) that show that the effects of driving, friction and 
circular error are additive. In this way we could deduce the expressions for 
these effects in a unified framework, without any reference to a phase diagram 
(Ref.~\onlinecite{RAW80}, Ch.~VII). In particular, the derivation of the 
escapement error in Eq.~(\ref{eq:dotpsi3}) runs smoothly and is free of ad 
hoc assumptions. 

We employ the method of Bogoliubov and Mitro\-pol\-ski \cite{BM61,BLV04}, 
which  involves averaging over a pendulum period. Time scales smaller than a 
period can no longer be resolved, but these are not important for the 
time keeping properties of the pendulum. The pleasant consequence is that the 
model equations are rather insensitive to fine structure in the driving torque 
$m(\varphi)$, which in turn allows us to use a simple model escapement. 
It is no longer necessary to resolve individual pendulum swings in small 
timesteps - we may now make timesteps of one pendulum period, resulting in a 
considerable time saving, and the possibility to make long simulations with 
all effects acting simultaneously. 

We studied two examples with the equation of motion (\ref{eq:ddt1}) and 
compared the results with those of the model equations (\ref{eq:dotl4}) and 
(\ref{eq:dotpsi4}): a sudden change in the driving torque and a permanently 
noisy driving. Satisfactory agreement was found. More complete models need to 
consider other disturbing factors; we only mention here that slow length 
changes $\delta\ell(t)$ of the pendulum produce an extra term $-\delta\ell/2
\ell_0$ on the right of Eqs.~(\ref{eq:dotpsi3}) and (\ref{eq:dotpsi4}), while 
Eqs.~(\ref{eq:dotl3}) and (\ref{eq:dotl4}) for the amplitude remain unchanged 
($\ell_0$ is the nominal pendulum length). If necessary new model equations 
may be derived for special escapements, see e.g. Eqs.~(\ref{eq:blow4}) and 
(\ref{eq:blow5}).

Now that the validation has been done, the road is open to fast simulations 
with the model equations (\ref{eq:dotl4}) and (\ref{eq:dotpsi4}). For example, 
the simulation of Fig.~\ref{fig:quasistat} comprises $\sim 110$ days for a 
$1\,$m pendulum. It took $10^4\,$s with the equation of motion (\ref{eq:ddt1}), 
but only $\sim 50\,$s with the model equations (\ref{eq:dotl4}) and 
(\ref{eq:dotpsi4}). Note that the reliability of such long-duration 
simulations is limited by the unknown influence of the second order terms. 
Evaluation of these second order terms seems to be an interesting topic now.                                       

It is evident from Eqs.~(\ref{eq:dotl3}) and (\ref{eq:dotpsi3}) that any fine 
structure of the type $\sum_{n\ne 1}(a_n\cos n\psi+b_n\sin n\psi)$ may be 
added to $m(\varphi)$ without affecting the time keeping. This opens the 
door to a wide variety of equivalent shapes for the driving torque $m$. But 
if the shape of $m$ is so unimportant, why did clockmakers bother to refine 
escapements? The answer must be that when parameters change and adapt to the 
environment, it is very difficult {\em not} to change the two quantities 
$\oint m\cos\psi\,\md\psi$ and $\oint m\sin\psi\,\md\psi$ that are crucial for 
the time keeping. The best strategy seems to be to reduce $\varepsilon$ as  
much as possible. 

A pendulum is the archetype harmonic oscillator, an omnipresent system in 
physics with widely different appearances.~\cite{BB05} Laboratory experiments 
with a driven, damped oscillator are therefore of permanent educational value. 
These need not be restricted to a pendulum. One might try for example a quartz 
crystal or any kind of electronic oscillator. But whatever the nature of the 
oscillator, the unifying concept of a driven, damped, restlessly swinging 
pendulum is likely to keep physicists and horologists spellbound forever.


\section*{Acknowledgements}
I am obliged to Dr. Matthijs Krijger for help with IDL and to Mr. Artur 
Pfeifer for help with the figures. I acknowledge a useful correspondence with 
Prof. Kees Grimbergen. And I thank the two unknown referees, whose expert 
remarks have led to a considerable improvement of the paper.


\appendix

\section{The circular error}
\label{sec:circer}
The circular error term follows from Eq.~(\ref{eq:dotpsi2}):
\begin{eqnarray} 
\!\!{\rm\ circ.\ error}\!\!&=&\!\!\oint
\{\lambda^{-1}\sin(\lambda\sin\psi)-\sin\psi\}\sin\psi\,\md\psi
\nonumber \\[2.mm]
&=&\!\frac{1}{\lambda}\,J_1(\lambda)-\frac{1}{2}\,\,
\simeq\,\!-\frac{1}{16}\,\lambda^2+\frac{1}{384}\,\lambda^4 
\,\cdot\cdot\,.
\qquad
\label{eq:ce3}
\end{eqnarray}
Here, $J_1$ is a Bessel function of the first kind. \cite{AS68} The essential 
steps in the computation above are (a) to use for $\sin(\lambda\sin\psi)$ 
relation 9.1.43 of Ref.~\onlinecite{AS68}, and then (b) the expansion 
$J_1(\lambda)=x(1-x^2/2+x^4/12-\cdots)$ with $x=\lambda/2$, cf. relation 
9.1.10 of Ref.~\onlinecite{AS68}.

We present Eq.~(\ref{eq:ce3}) here because it differs from the usual 
expansions (e.g. Ref.~\onlinecite{RAW80}, p.~51 and Ref.~\onlinecite{NO86}, 
Eq.~(8). This is because Eq.~(\ref{eq:ce3}) specifies the effect of circular 
error on the rate, not on the period. 


\section{Back-of-the-envelope derivation of equations (\ref{eq:dotl3}) and 
(\ref{eq:dotpsi3})}
\label{sec:appeqs}
The power fed into the pendulum is torque $\times$ angular speed $=m\dot
\varphi=m\lambda\cos\psi$ (use Eq.~(\ref{eq:deflabpsi}) for $\dot\varphi$; 
gravity does on average no work). But this is also the time derivative of the 
total energy, so $\md(\hlf\lambda^2)/\md t=m\lambda\cos\psi$ or $\dot\lambda=
m\cos\psi$. Eq.~(\ref{eq:dotl3}) follows by adding friction and averaging. 

Suppose $m$ is a single impulse. At that instant $\dot\varphi$ gets a boost 
while $\varphi=\lambda\sin\psi$ is constant, so $0=\md(\lambda\sin\psi)/
\md t=\dot\lambda\sin\psi+\lambda\dot\psi\cos\psi$. Solve for $\dot\psi$ 
and insert $\dot\lambda=m\cos\psi$ to obtain $\dot\psi=-(m/\lambda)\sin\psi$. 
This is the core of Eq.~(\ref{eq:dotpsi3}). Add all impulses $m_i$ composing 
$m$ in one period: $\delta^\prime\psi=\allowbreak\sum\delta\psi_i=\allowbreak
-(1/\lambda)\sum m_i\sin\psi_i\,\delta t=\allowbreak-(1/\lambda)\int m\sin
\psi\,\md t$ ($\lambda$ is constant over a period). Denote the mean time 
derivative of $\psi$ again by $\dot\psi$, then $\dot\psi\simeq \delta^\prime
\psi/2\pi\simeq-(1/\lambda)\int m\sin\psi\,\md t/2\pi\,\simeq-(1/\lambda)
\oint m\sin\psi\,\md\psi$ (since $\md\psi/\md t=1+{\rm small\ terms}$, so $\md 
t\simeq\md\psi$). Finally, add the nominal rate $\dot\psi=1$, and the circular 
error $-\lambda^2/16$. 

It is hoped that these informal arguments help to clarify those of 
Ref.~\onlinecite{RAW80}, Ch.~7 and Woodward. \cite{W76}
\\

\section{A single-impulse escapement}  
\label{sec:appblow}
The model escapement, Fig.~\ref{fig:modesc}, that we have used sofar 
interferes continuously with the pendulum motion. Here we compare it with an 
escapement of opposite type, that delivers in each period a single impulse 
at a fixed pendulum angle $\varphi_1$:  
\begin{equation}
m(\varphi)=e\,\delta(\varphi-\varphi_1)\ .
\label{eq:blow1}
\end{equation}
The function $\delta(x)$ is basically a sharp peak at $x=0$. Width and height 
are inversely proportional such that $\int\delta(x-x_0)\,\md x=1$, provided 
$x_0\in I,$ the integration interval. The width of $\delta(x)$ is supposed to 
be much smaller than all fine structure in the problem at hand, whence $\int 
f(x)\delta(x-x_0)\,\md x\simeq f(x_0)$ if $x_0\in I$, otherwise zero. 

The energy transferred to the pendulum in one period is $\int m\,\md\varphi=
e>0$. It follows that $\md\varphi>0$ at $\varphi_1$, i.e. $\varphi$ is 
increasing. Since $\varphi_1=\lambda\sin\psi_1$ according to 
Eq.~(\ref{eq:deflabpsi}), the phase $\psi_1$ at the impulse is in the 1st or 
4th quadrant. We compute $a$ and $b$ given by Eqs. (\ref{eq:ca1}) and 
(\ref{eq:cb1}):
\begin{equation}
b=\oint m\sin\psi\,\md\psi=\frac{e}{2\pi}\int_{-\pi}^\pi
\delta(\varphi-\varphi_1)\sin\psi\,\md\psi\ .
\label{eq:blow2}
\end{equation}
Here we hit a technicality: to be able to apply the $\delta$-function recipe 
we transform the integration over $\psi$ to one over $\varphi$. From $\varphi=
\lambda\sin\psi$ and $\lambda\simeq$ constant in a period, we infer $\md\psi=
(\md\varphi/\md\psi)^{-1}\md\varphi=\allowbreak(\lambda\cos\psi)^{-1}
\md\varphi$, so 
\begin{equation}
b=\frac{e}{2\pi\lambda}\int_{-\lambda}^\lambda
\tan\psi\;\delta(\varphi-\varphi_1)\,\md\varphi
=\frac{e}{2\pi\lambda}\,\tan\psi_1\ .
\label{eq:blow3}
\end{equation}
A similar computation yields $a=e/(2\pi\lambda)$. To obtain the new model 
equations for $\lambda$ and $\psi$ we repeat the calculation of 
Sec.~\ref{sec:modesc}, resulting in:
\begin{eqnarray}
\dot\lambda\!&=&\!-\varepsilon\lambda+\frac{e}{2\pi\lambda}\ ,
\label{eq:blow4} \\
\dot\psi-1\!&=&\!-\frac{\lambda^2}{16}-\frac{e}{2\pi\lambda^2}\,\tan\psi_1\ .
\label{eq:blow5}
\end{eqnarray}
These equations may be used to study the performance of a single-impulse 
escapement. We briefly look at the case of stationary operation. The amplitude 
is then $\lambda=(e/2\pi\varepsilon)^{1/2}$ and the rate is:
\begin{equation}
\dot\psi-1=-\frac{e}{32\pi\varepsilon}-\varepsilon\tan\psi_1\ .
\label{eq:blow6}
\end{equation}
The last term is the escapement error, which now either increases the rate 
($\psi_1<0$, impulse before a zero), or reduces it ($\psi_1>0$, impulse after 
a zero), cf. Ref.~\onlinecite{RAW80}, p. 130. Note that $\varphi_1=\lambda\sin
\psi_1$ implies $\tan\psi_1=\varphi_1(\lambda^2-\varphi_1^2)^{-1/2}$, and just 
as in Eq. (\ref{eq:dotpsi4}), we see the appearance of the square root functions 
in Ch.~VII of Ref.~\onlinecite{RAW80}.


\section{Numerical issues}
\label{sec:appnum}
We rewrite Eq.~(\ref{eq:ddt1}) as a first order ODE in the variable $(\varphi,
\dot\varphi)$ which we solve with the IDL routine RK4. To model the escapement 
the code finds out, at each timestep, in which quadrant $\varphi$ lies (from 
the signs of $\varphi$ and $\dot\varphi$). From the values of $\varphi$ and 
$\varphi_0$ the code then decides between $m=+\mu$ and $m=-\mu$, according to 
Fig.~\ref{fig:modesc}. 

The simulation delivers pendulum angles $\varphi_k$ on an equidistant grid 
$t_k$, stepsize $dt$. To measure the rate with Eq.~(\ref{eq:ratemeas}) we need 
the $-+$ zero crossing times of $\varphi$ with a precision much better than 
$\sim\varepsilon$. This follows from Eqs.~(\ref{eq:ratest}) and 
(\ref{eq:ratemeas}). It is therefore necessary to do the whole (IDL) simulation 
in double precision. The timestep $k$ enclosing a zero obeys $\varphi_{k-1}<0$ 
and $\varphi_k>0$. An approximate zero crossing time is found by linear 
interpolation: $t_{\rm zero}=t_{k-1}+dt\cdot\varphi_{k-1}/(\varphi_{k-1}-
\varphi_k$). This gave an acceptable accuracy, though we do observe some 
numerical noise, for example in Fig.~\ref{fig:jump}, top panel.


\end{document}